\documentclass[12pt]{article}
\usepackage[dvips,bookmarks]{hyperref}
\hypersetup{colorlinks,bookmarksopen,bookmarksnumbered,citecolor=blue,
	pdfstartview=FitH}
\def\hhref#1{\href{http://arxiv.org/abs/hep-th/#1}{hep-th/#1}}
\def\mhref#1{\href{mailto:#1}{#1}}
\begin{document}
\newcommand{\p}{\partial}
\newcommand{\hp}{\hat{\p}}
\newcommand{\ov}{\overline}
\newcommand{\da}{^{\dagger}}
\newcommand{\w}{\wedge}
\newcommand{\al}{\alpha}
\newcommand{\bb}{\beta}
\newcommand{\ga}{\gamma}
\newcommand{\te}{\theta}
\newcommand{\de}{\delta}
\newcommand{\et}{\tilde{e}}
\newcommand{\ze}{\zeta}
\newcommand{\s}{\sigma}
\newcommand{\e}{\epsilon}
\newcommand{\om}{\omega}
\newcommand{\Om}{\Omega}
\newcommand{\la}{\lambda}
\newcommand{\La}{\Lambda}
\newcommand{\n}{\nabla}
\newcommand{\hn}{\hat{\nabla}}
\newcommand{\hph}{\hat{\phi}}
\newcommand{\ah}{\hat{a}}
\newcommand{\bh}{\hat{b}}
\newcommand{\ch}{\hat{c}}
\newcommand{\eh}{\hat{e}}
\newcommand{\ph}{\hat{p}}
\newcommand{\qh}{\hat{q}}
\newcommand{\mh}{\hat{m}}
\newcommand{\nh}{\hat{n}}
\newcommand{\as}{\breve{a}}
\newcommand{\bs}{\breve{b}}
\newcommand{\cs}{\breve{c}}
\newcommand{\ds}{\breve{d}}
\newcommand{\es}{\breve{e}}
\newcommand{\ms}{\breve{m}}
\newcommand{\ns}{\breve{n}}
\newcommand{\ps}{\breve{p}}
\newcommand{\ad}{\dot{a}}
\newcommand{\bd}{\dot{b}}
\newcommand{\gd}{\dot{c}}
\newcommand{\dd}{\dot{\delta}}
\newcommand{\ed}{\dot{\eta}}
\newcommand{\zd}{\dot{\zeta}}
\newcommand{\md}{\dot{m}}
\newcommand{\nd}{\dot{n}}
\newcommand{\az}{\grave{a}}
\newcommand{\bz}{\grave{b}}
\newcommand{\nz}{\grave{n}}
\newcommand{\mz}{\grave{m}}
\newcommand{\tb}{\overline{\theta}}
\newcommand{\ti}{\widetilde}

\newcommand{\2}{\textstyle{1\over 2}}
\newcommand{\3}{\frac{1}{3}}
\newcommand{\4}{\frac{1}{4}}
\newcommand{\8}{\frac{1}{8}}
\newcommand{\6}{\frac{1}{16}}
\newcommand{\ra}{\rightarrow}
\newcommand{\Ra}{\Rightarrow}
\newcommand{\im}{\Longleftrightarrow}
\newcommand{\hs}{\hspace{5mm}}
\newcommand{\x}{\star}
\newcommand{\Delt}{\p^{\star}}

\thispagestyle{empty}
{\bf 29th October, 2002} \hspace{\fill}
{\bf YITP-SB-02-60}

\vspace{1cm}
\begin{center}{\Large{\bf SYMMETRY BREAKING OF GAUGE THEORIES\\

\vspace{3mm}
 VIA INTERNAL SPACE DYNAMICS}}\\
\vspace{1cm}
{\large{\bf T. Biswas\footnote{\mhref{tirtho@insti.physics.sunysb.edu} }}}\\
\vspace{5mm}
{\small C.N. Yang Institute of Theoretical Physics\\
Department of Physics and Astronomy\\
State University of New York at Stony Brook\\
Stony Brook, New York 11794-3840}
\end{center}

\begin{abstract}
In this paper, I explain how gauge symmetry can be broken in a geometric way, \`{a} la Kaluza-Klein. In higher dimensional gravitational theories, one usually considers the  extra dimensions to be ``frozen'' in time. However, the  internal manifold is actually a dynamic entity. For example, its  metric can change even if one expects its topological properties to be invariant. It is conceivable  then, that at an earlier epoch the internal manifold made a geometric transition from say a maximally symmetric metric space to a less symmetric one. We know in a Kaluza-Klein reduction scheme, the massless gauge bosons are associated with the Killing vectors of the internal manifold.  After the  transition of the internal manifold, the gauge bosons associated with the broken Killing isometries will pick up a mass thereby breaking the gauge invariance partially. In this paper, I explore this idea, work out the mass of broken gauge bosons  for some simple examples, and also point out how a mechanism similar to that of Higgs may be at work.
\end{abstract}

\newpage
\setcounter{page}{1}

\section{{\bf   INTRODUCTION}}

Symmetry and symmetry-breaking are two key ingredients in modern high energy physics. While symmetry brings with it unity and simplicity, one often also needs  to break it by some mechanism to account for the observations in our universe; the Standard Model being a notable example. Till now various such mechanisms are known: Higgs mechanism \cite{higgs} or spontaneous symmetry breaking which was very successfully applied to the Standard Model, dynamic symmetry breaking via vacuum condensation \cite{coleman}, etc. The earliest idea of using geometry of extra dimensions to break gauge symmetry can be found  in the context of ``dimensional reduction by isometries'' \cite{cho}, which is however fundamentally different from the Kaluza-Klein scenario. In this paper, I propose a new geometric mechanism for breaking gauge symmetry in the context of Kaluza-Klein reduction schemes. 

Higher dimensional, i.e. greater than 4, (super) gravitational theories are beautiful in that they combine (four-dimensional) gravity with gauge interactions through Kaluza-Klein reduction  schemes (see for example \cite{kaluza,duff} for details) in a very geometric way. One considers the vacuum to be a product of a four-dimensional vacuum manifold (Minkowski, deSitter or anti deSitter), and an (usually compact) internal manifold with matching scalar curvature constants. Four-dimensional physics then arises as fluctuations around this vacuum. For example, if one looks at the massless modes, which are  important for describing low energy physics of the higher dimensional metric, then one finds a  graviton (in the four-dimensional sector of the metric) and  gauge bosons (appearing in the off-diagonal part of the metric) associated with the Killing vectors of the ``frozen'' internal manifold. After the Kaluza-Klein reduction of the higher dimensional gravitational action, one obtains a covariant and gauge invariant action of gravity coupled with gauge bosons among other things. However, as I mentioned earlier, we also need a mechanism to break the gauge symmetry (at least partially) to make it compatible with the experimental observations, and what can be better if we also find a geometric origin behind this symmetry breaking! 

Such a geometric mechanism emerges naturally when one realises that the internal manifold is dynamic in nature, more prominently at earlier times. It is possible then that the internal manifold started out as say, a maximally symmetric metric space but made a transition to a less symmetric metric space.  Such a vacuum to vacuum (asymptotically) transition can for example, occur along a classical path (if such a path is available), or through quantum tunneling etc. At any case after the transition, some of the original Killing symmetries will no longer remain Killing vectors. Let us name the initial and final isometry group of the internal manifold as $G$ and $H$ respectively ($H \subset G$). The gauge bosons originally associated with the broken Killing symmetries will now pick up a mass term in the action thereby breaking the gauge symmetry group from $G$ to $H$. 

One can indeed calculate the masses of the broken gauge bosons, and the various mass ratios would be a precise prediction coming out of these models. Further given a  $H \subset G$, it is not always possible to find an Einstein metric on the given internal manifold with Killing symmetry group $H$. Thus this model also narrows down the list of subgroups that a gauge group can be broken down to. In this paper, I concentrate on the simple case when the internal manifold is a Lie group ($L$) itself, and the maximally isometric group $L_{left} \otimes L_{right}$ is broken down to $M_{left} \otimes L_{right}$, $M \subset L$ and in particular work out the details when $L=SU(3)$, which is really the simplest example where such a  geometric transition is possible. Further, in this paper I only concern myself with pure gravity. Since the analysis relies completely on ``geo-metric'' ideas, it should be straight-forward to embed it in larger frameworks of unified  theories of gravity like the String and supergravity (SUGRA) theories. It should also be possible to generalise the internal manifold to coset spaces; and in principle to nonhomogeneous spaces which may be more relevant to the String/SUGRA compactifications. A crucial consideration when dealing with these different vacua of the  internal manifold is the issue of stability \cite{stability}.  It will be specially interesting to find cases where the $M_{left} \otimes L_{right}$-invariant  metric is stable but the $L_{left} \otimes L_{right}$-invariant  one is not, making the geometric transition a very viable prospect. 

A natural question that arises in such a scenario is what happened to the counting of states. With each broken Killing symmetry, the number of physical states increases by one because the vector becomes massive from massless. The compensation occurs in a manner very similar to that of the Higgs' mechanism. The ``freed'' gauge parameters kill some of the scalars (in the adjoint representation of the original gauge group) that appear as fluctuations in the metric of the internal manifold (see section 4 for details). 

I first review some facts about Lie group geometry (see for example \cite{liegroup}) in section 1, and in particular talk about metrics in $SU(2)$ and $SU(3)$. In section 2, I provide a qualitative understanding of the dynamics of the transition and obtain the ``mass matrix'' for the broken gauge bosons when the internal manifold is a Lie group. As an illustration, I also compute explicitly the mass matrix when the gauge group is broken from $SU(3)_{left}\otimes SU(3)_{right}$ to $SO(3)_{left}\otimes SU(3)_{right}$, the internal manifold being  $SU(3)$. In section 3, I elucidate on the Higgs-like mechanism that helps preserve the number of physical states before and after the geometric transition. I conclude by summarizing and making some remarks about possible future research.
  
\section{{\bf  LIE GROUPS AS INTERNAL MANIFOLD}}
{\bf Geometry of Lie groups:} A Lie group element $g$ can be parametrized as 
\begin{equation}
g=exp(\breve{\la}^{\as}(y^{\ms})T_{\as})\ \e\ G
\end{equation}
where $T_{\as}\ \e\ \cal{G}$, the Lie algebra corresponding to the Lie group $G$ and $\breve{\la}^{\as}(y^{\ms})$ are some given functions of the coordinates $y^{\ms}$ charting the Lie group manifold. The Lie group generators $T_{\as}$ satisfy the usual commutation relations:
\begin{equation}
[T_{\as},T_{\bs}]=C_{\as\bs}{}^{\cs}T_{\cs}
\end{equation}
where $C_{\as\bs}{}^{\cs}$ are the structure constants of the Lie group. With each of the generators $T_{\as}$, one can associate a left and a right invariant vector field $e_{\as}$ and $\et_{\as}$ respectively. Both sets $\{e_{\as}\}$ and $\{\et_{\as}\}$ can serve as vielbeins or local basis vector fields for the tangent space of the Lie group. They are defined via the following relations
\begin{equation}
e_{\as}\equiv e_{\as}{}^{\ms}\p_{\ms};\ \et_{\as}\equiv \et_{\as}{}^{\ms}\p_{\ms}
\end{equation}
\begin{equation}
e_{\as}{}^{\ms}\equiv (e_{\ms}{}^{\as})^{-1};\ \et_{\as}{}^{\ms}\equiv (\et_{\ms}{}^{\as})^{-1}
\end{equation}
and
\begin{equation}
g^{-1}\p_{\ms}g=e_{\ms}{}^{\as}T_{\as} ;\ (\p_{\ms}g)g^{-1}=\et_{\ms}{}^{\as}T_{\as}
\end{equation}
These two reference frames are related by a local Lorentz transformation
\begin{equation}
\et_{\as}=D_{\as}{}^{\bs}(g)e_{\bs}
\end{equation}
where $D_{\as}{}^{\bs}(g)$ is the adjoint representation of $G$. In the subsequent discussion we will choose $\{e_{\as}\}$ as the local frame of reference. In this frame, a general metric on $G$ looks like
$$ g_{\as\bs}=g_{\as\bs}(y^{\ms})$$
However, we are interested in metrics with special symmetry properties. It can be shown that in general the isometry group of a metric will be $H_L\otimes K_R$, where $H,K\subseteq G$. In particular \begin{equation}
K=G \Ra g_{\as\bs}=\mbox{constants}
\end{equation}
We will be principally concerned with such right invariant metrics. These metrics are invariant under the right invariant vector fields $\{\et_{\as}\}$, but not in general under the left invariant vector fields. This follows readily from the commutation relations between them:
\begin{equation}
[e_{\as},e_{\bs}]=C_{\as\bs}{}^{\cs}e_{\cs}\ ;\ [\et_{\as},\et_{\bs}]=-C_{\as\bs}{}^{\cs}\et_{\cs}\ ;\ [\et_{\as},e_{\bs}]=0
\end{equation}
If we want the metric to be further invariant under say $H_L$, then it has to satisfy
\begin{equation}
g_{\as\bs}=D_{\as}{}^{\cs}(h)D_{\bs}{}^{\ds}(h)g_{\cs\ds}\ \forall\  h\ \e \ H
\end{equation}
If we suitably choose our generators $\{T_{\as}\}=\{T_{\az},T_{\ad}\}$ such that $\{T_{\ad}\}$ span $\cal{H}$, then the Killing vectors of this $H_L\otimes G_R$ (right) invariant metric will be the $\{\et_{\as}\}$'s and the $\{e_{\ad}\}$'s. I am using the symbols $\breve{},\ \dot{}$, and $\grave{}$ to indicate quantities associated with $G$, $H$ and the coset space $G/H$ respectively. A special case of the right invariant metric is the bivariant metric when $H=G$, i.e. it has the maximal isometry, and is invariant under both $\{\et_{\as}\}$ and $\{e_{\as}\}$'s. The Killing metric given by
\begin{equation}
g_{\as\bs}=-C_{\as\cs}{}^{\ds}C_{\bs\ds}{}^{\cs}
\end{equation}
is an example of such a metric. Further, the Killing metric satisfies Einstein's field equations,  and hence is consistent with its usual identification as Kaluza-Klein vacuum. However, the Killing metric is not the only  right invariant metric which is Einstein, for some specific $H_L$'s and specific parameter values we can  hope to find other right invariant Einstein metrics, and hence a Kaluza-Klein vacuum.
\vspace{5mm}
\\ 
{\bf SU(2) and SU(3):} For the 3-dimensional Lie group manifold $SU(2)$, the most general (modulo local Lorentz transformations) right invariant metric that one can write down looks like 
\begin{equation}
g_{\as\bs}=\left( \begin{array}{ccc}
g_{11} & 0& 0\\
0 & g_{22}& 0\\
0& 0&g_{33}
\end{array} \right)
\end{equation}
One can now try to solve Einstein's vacuum field equations:
\begin{equation}
R_{\as\bs}=\breve{\la}g_{\as\bs}
\end{equation}
where $\breve{\la}$ is related to the cosmological constant. It turns out that the only solution corresponds to
\begin{equation}
g_{11}=g_{22}=g_{33}
\end{equation}
which is proportional to the bivariant Killing metric. Thus there is no Einstein $H_L\otimes SU(2)_R$ metric,  and consequently we cannot break the gauge group $SU(2)_L\otimes SU(2)_R$ to $H_L\otimes SU(2)_R$ with $H\subset SU(2)$.

For $SU(3)$ however, we do have more than one right invariant Einstein metric. A convenient and simple way to find such metrics is to start with the Killing metric and then scale the metric sector corresponding to the subgroup $H$ that we want to preserve by a parameter which I will refer to as $T^2$:
\begin{equation}
g^S_{\as\bs}=\left( \begin{array}{cc}
g^K_{\az\bz} & 0\\
0& T^2g^K_{\ad\bd}
\end{array} \right)
\end{equation}
Here $g^K$ and $g^S$ are the Killing and scaled metric respectively. Now for specific values of this parameter one may  find Einstein metrics. Applying the ansatz (2.14) to (2.12) for $SU(3)$ one can find a vacuum metric  which is invariant under $SO(3)_L\otimes SU(3)_R$,  but none which is invariant under  $(U(1)\otimes U(1))_L\otimes SU(3)_R$ or  $(SU(2)\otimes U(1))_L\otimes SU(3)_R$. 

To see this in more detail let us first label the generators.
$$
T_{10}=\left( \begin{array}{ccc}
1& 0&0\\
0& -1&0\\
0& 0& 0
\end{array} \right)\ ;\ 
T_{20}=\left( \begin{array}{ccc}
0& 0&0\\
0& 1&0\\
0& 0& -1
\end{array} \right)
$$
$$
T_{1A}=\left( \begin{array}{ccc}
0& 1&0\\
1& 0&0\\
0& 0& 0
\end{array} \right)\ ;\ 
T_{2A}=\left( \begin{array}{ccc}
0& 0&0\\
0& 0&1\\
0& 1& 0
\end{array} \right)\ ;\ 
T_{3A}=\left( \begin{array}{ccc}
0& 0&1\\
0& 0&0\\
1& 0& 0
\end{array} \right)
$$
\begin{equation}
T_{1R}=\left( \begin{array}{ccc}
0& i&0\\
-i& 0&0\\
0& 0& 0
\end{array} \right)\ ;\ 
T_{2R}=\left( \begin{array}{ccc}
0& 0&0\\
0& 0&i\\
0& -i& 0
\end{array} \right)\ ;\ 
T_{3R}=\left( \begin{array}{ccc}
0& 0&-i\\
0& 0&0\\
i& 0& 0
\end{array} \right)
\end{equation}
Plugging the structure constants corresponding to the generators (2.15) in (2.10) one computes the Killing metric:
\begin{equation}
g^K_{\as\bs}=12\left( \begin{array}{cccccccc}
1& -\2&0&0&0&0&0&0\\
-\2& 1&0&0&0&0&0&0\\
0& 0  &1&0&0&0&0&0\\
0& 0  &0&1&0&0&0&0\\
0& 0  &0&0&1&0&0&0\\
0& 0  &0&0&0&1&0&0\\
0& 0  &0&0&0&0&1&0\\
0& 0  &0&0&0&0&0&1
\end{array} \right)\ 
\end{equation}
Here the ordering of the rows and columns is the same as the order in which the generators are written above. The ansatz for a $SO(3)_L$ (generated by the $T_R$'s) invariant metric then looks like
\begin{equation}
g^{SO(3)}_{\as\bs}=12\left( \begin{array}{cccccccc}
1& -\2&0&0&0&0&0&0\\
-\2& 1&0&0&0&0&0&0\\
0& 0  &1&0&0&0&0&0\\
0& 0  &0&1&0&0&0&0\\
0& 0  &0&0&1&0&0&0\\
0& 0  &0&0&0&T^2&0&0\\
0& 0  &0&0&0&0&T^2&0\\
0& 0  &0&0&0&0&0&T^2
\end{array} \right)\ 
\end{equation}
and plugging (2.17) in (2.12), one finds (2.17) is Einstein for 
\begin{equation}
T^2=1,\ \breve{\la}=\4;\mbox{ and } T^2=\frac{1}{11},\ \breve{\la}=\frac{21}{44}
\end{equation}
While the former corresponds to the Killing metric, the latter is a $SO(3)_L\otimes SU(3)_R$ invariant metric.

In a similar manner, if one wants to obtain an $(U(1)\otimes U(1))_L$ (generated by the $T_0$'s) invariant metric, the ansatz then looks like
\begin{equation}
g^{U(1)\otimes U(1)}_{\as\bs}=12\left( \begin{array}{cccccccc}
T^2& -\2 T^2&0&0&0&0&0&0\\
-\2 T^2& T^2&0&0&0&0&0&0\\
0& 0  &1&0&0&0&0&0\\
0& 0  &0&1&0&0&0&0\\
0& 0  &0&0&1&0&0&0\\
0& 0  &0&0&0&1&0&0\\
0& 0  &0&0&0&0&1&0\\
0& 0  &0&0&0&0&0&1
\end{array} \right)\ 
\end{equation}
which  satisfies Einstein's field equations for
\begin{equation}
T^2=1,\ \breve{\la}=\4;\mbox{ and } T^2=-\frac{5}{3},\ \breve{\la}=-\frac{5}{12}
\end{equation}
The latter corresponds to an $(U(1)\otimes U(1))_L\otimes SU(3)_R$ invariant metric, but it has a different signature and hence is not a ``proper'' vacuum metric on $SU(3)$. 

The same thing happens when we try to obtain an  $U(1)_L\otimes SU(3)_R$ invariant Einstein metric. Using a similar ansatz one finds that there are no nonsingular Einstein $(SU(2)\otimes U(1))_L\otimes SU(3)_R$  invariant metric. In fact, the only choices for a $H_L\otimes G_R$ invariant vacuum metric are $H=SO(3)$ or $SU(3)$.

\section{{\bf   SYMMETRY BREAKING AND MASS MATRIX}}
{\bf Kaluza-Klein Vacua:} We consider a $D$ dimensional space-time manifold with 4 ordinary (large) space-time dimensions, labeled by $x^{m}$  and $D-4$ extra compact dimensions forming the ``internal manifold'' charted by $y^{\ms}$. $\{x^{\mh}\}=\{x^m,y^{\ms}\}$. The vielbein and its inverse, corresponding to the Kaluza-Klein vacuum is then given by
\begin{equation}
\eh_{\mh}{}^{\ah}=\left( \begin{array}{cc}
e_m{}^{a} & 0\\
0 &\es_{\ms}{}^{\as}
\end{array} \right);
\hspace{5mm}
\eh_{\ah}{}^{\mh}=\left( \begin{array}{cc}
e_{a}{}^{m} & 0\\
0 &\es_{\as}{}^{\ms}
\end{array} \right)
\end{equation}
with the Einstein metric  
\begin{equation}
\hat{g}_{\ah\bh}=\left( \begin{array}{cc}
g_{ab} & 0\\
0 & g_{\as\bs}
\end{array} \right)
\end{equation}
Here the four-dimensional part of the vielbein corresponds to either the  Minkowski, or the  deSitter spaces according to the sign of the scalar curvature constant of the internal manifold.  For (3.1-3.2) to be a vacuum, $\{\es_{\as}{}^{\ms},g_{\as\bs}\}$ also need to satisfy the $D-4$ dimensional Einstein's field equations for the internal manifold. 

Historically, although seeds of nonabelian generalization of Kaluza-Klein reduction scheme can be seen in as early as Klein's and Pauli's works (see \cite{dawn,kaluza} for details), it was Kerner \cite{kerner} who first performed a complete nonabelian analysis by considering  Lie groups as internal manifolds. We will also specialize to the case when the internal manifold is a (semisimple) Lie group $G$. We already  know that there exists one Einstein vacuum in the form of the Killing metric ($g^K_{\as\bs}$) which is invariant under $G_L\otimes G_R$, and we will also assume that there is another Einstein metric $g^S_{\as\bs}$ which is  invariant under $H_L\otimes G_R$. We can choose the Lie group generators $\{T_{\as}\}=\{T_{\az},T_{\ad}\}$ (as discussed in the previous section) so that $\{T_{\ad}\}$ generates $\cal{H}$. We can then parameterise the group element as 
$$g=exp(\grave{\la}^{\az}(y^{\mz})T_{\az}) exp(\dot{\la}^{\ad}(y^{\md})T_{\ad})$$
Let us now consider the situation when the internal manifold makes a transition from $g^K_{\as\bs}$ to $g^S_{\as\bs}$. Now, all the left and the right invariant vector fields are  Killing vectors for the Killing metric, but the Killing vectors of the scaled metric are given by the right invariant vector fields 
$$\ti{\es}_{\as}=\ti{\es}_{\as}{}^{\ms}\p_{\ms}$$
and the left invariant vector fields 
\begin{equation}
\es_{\ad}=\es_{\ad}{}^{\md}\p_{\md}
\end{equation}
Thus the $\es_{\az}$'s constitute the broken Killing symmetry generators.
 \vspace{5mm}
\\ 
{\bf Transition and Reduction:} We are now ready to investigate the fluctuations around the Kaluza Klein vacuum. We will be principally interested in the massless modes as they are the most important ones in the low energy description of physics. Further, although these ideas should be applied ultimately to supergravity (SUGRA) theories (see the subsection ``Dynamical Considerations and Supergravity'' for detail), for simplicity we will only concern ourselves  with the massless modes coming from the $D$ dimensional metric here. It is  well known what these massless modes are: It consists of the spin 2 graviton (fluctuations around the four-dimensional sector of the vacuum vielbein), and spin 1 gauge bosons associated with the Killing vectors of the internal manifold (occurring as fluctuations around the off diagonal part of the vielbein)\footnote{Although the dilaton is supposed to become massive in the full quantum theory, it does play an important role in the low energy physics. However, since presently we are only interested in the fluctuations corresponding to the broken symmetries we can ignore it in our analysis.}. Now consider what happens to these modes if the internal manifold itself makes a transition from a maximally isometric space ($G_L\otimes G_R$ invariant metric) to a less symmetric space ($H_L\otimes G_R$ invariant, $H\subset G$).  Such a geometric transition can take place at an earlier cosmic time when the energy scales involved suggest a truely dynamic internal manifold. Of course, such a vacuum to vacuum geometric transition  cannot be understood  perturbatively, and to analyse it comprehensively is indeed a difficult task. In this paper, I will only give a brief outline of what such a program will entail (see ``Dynamical Considerations and Supergravity'') and leave the details for further research. We now focus on the consequences of such a transition, in particular on the massless modes. 

Nothing happens to gravity, but clearly the gauge bosons associated with the broken  Killing vectors will now no longer be massless. The initial $G_L\otimes G_R$ nonabelian gauge theory, consisting of  $\ti{A}^{\as}_m$ and $A^{\as}_m$, associated with $\ti{\es}_{\as}$'s and $\es_{\as}$'s respectively, is now partially broken  to a $H_L\otimes G_R$ gauge theory, comprising of the gauge fields $\ti{A}^{\as}_m$ and $A^{\ad}_m$, while $A^{\az}_m$'s pick up  mass terms. Our aim now is to compute the ``mass-matrix'' for these broken gauge fields.

Since we agree that the graviton $g_{mn}$, and the gauge bosons $\ti{A}^{\as}_m$ associated with the right invariant  vectors remain massless after the transition, we can focus only on the fluctuations of the vielbein due to the left invariant vector fields. 
\begin{equation}
\eh_{\mh}{}^{\ah}=\left( \begin{array}{cc}
e_m{}^{a}(x) & -A_m^{\as}(x)\\
0 &\es_{\ms}{}^{\as}(y)
\end{array} \right);
\hspace{5mm}
\eh_{\ah}{}^{\mh}=\left( \begin{array}{cc}
e_{a}{}^{m}(x) & \ e_a{}^n(x)A_n^{\bs}(x)\es_{\bs}{}^{\ms}(y)\\
0 &\ \es_{\as}{}^{\ms}(y)
\end{array} \right)
\end{equation}
The form of the vielbein is not to be treated as an ansatz, which has to then satisfy several consistency conditions\footnote{For example it was shown in \cite{consistency} that in an ansatz for the vielbein of the form (3.4), we cannot have gauge fields associated with both the left and the right invariant vector fields in the off-diagonal sector. However, undoubtedly they both correspond to massless modes suggesting a $G_L\otimes G_R$ gauge symmetry in the low energy effective theory (once all the modes necessary for a consistent truncation are taken into account) which is broken, after the transition to $H_L\otimes G_R$. If one is still worried about the consistency requirements, this scheme can be thought of as breaking a $G_L$ gauge theory to $H_L$.}, but rather as containing the relevant fluctuations to describe the low energy fluctuations. For example, this will enable us to compute the gauge boson mass generated through this transition.

It is a straight-forward exercise to compute $\sqrt{-\hat{g}}\hat{R}$. One obtains
\begin{equation}
\hat{R}=R+\breve{R}-\frac{1}{4}(F_{ab}^{\cs}F^{ab}_{\cs}+A^{a\bs}A_a^{\cs}M_{\bs\cs})
\end{equation}
where
\begin{equation}
F_{ab}^{\cs}=e_a{}^me_b{}^n[\p_{[m}A^{\cs}_{n]}+C_{\as\bs}{}^{\cs}A_m^{\as}A_n^{\bs}]
\end{equation}
and
\begin{equation}
M_{\as\bs}=2[C_{\as\cs}{}^{\ds}C_{\bs\ds}{}^{\cs}+C_{\as\cs}{}^{\ds}C_{\bs}{}^{\cs}{}_{\ds}]
\end{equation}
while 
\begin{equation}
\sqrt{-\hat{g}}=\hat{e}=e.\es
\end{equation}
Thus we have
\begin{equation}
\sqrt{-\hat{g}}R=e.\es[R+\breve{R}-\frac{1}{4}(F_{ab}^{\cs}F^{ab}_{\cs}+A^{a\bs}A_a^{\cs}M_{\bs\cs})]
\end{equation}
We note that both $R$ and $\breve{R}$ are constants; since both the four-dimensional and the internal manifold are Einstein and hence have constant scalar curvatures. Thus after integrating out the extra compact dimensions, we have the effective action for the gauge fields:
\begin{equation}
S_{gauge,eff}=-\frac{1}{4}\mbox{Vol}_G\int dx\ e(F_{ab}^{\cs}F^{ab}_{\cs}+A^{a\bs}A_a^{\cs}M_{\bs\cs})
\end{equation}\\
Actually, as is clear from the action, $M_{\as\bs}$ really represents the mass square. For the maximally symmetric internal manifold it vanishes, whereas the broken sector is non-vanishing for the scaled metric. 
\vspace{5mm}
\\ 
{\bf The example of SU(3):} We had previously seen that the only Einstein right invariant metric on $SU(2)$ is also invariant under $SU(2)_L$. Hence a geometric symmetry breaking to $U(1)_L$ cannot occur. $SU(3)$ is the ``simplest'' group manifold which can exibit a geometric transition. In section 2, we obtained some right invariant metrics on $SU(3)$ which were Einstein, out of which only the Killing metric and the $SO(3)_L$ invariant metric  can become consistent Kaluza-Klein vacua. Thus the ``geometric transition'' mechanism predicts that if we want to break the  gauge group $SU(3)_L\otimes SU(3)_R$ to $H_L\otimes SU(3)_R$, $H\subset SU(3)$, then $H$ has to be $SO(3)$.

As expected the mass matrix for the Killing metric vanishes, but for the $SO(3)$-scaled metric it looks like
\begin{equation}
M_{\as\bs}=\frac{1200}{11}\left( \begin{array}{cccccccc}
1& -\2&0&0&0&0&0&0\\
-\2& 1&0&0&0&0&0&0\\
0& 0  &1&0&0&0&0&0\\
0& 0  &0&1&0&0&0&0\\
0& 0  &0&0&1&0&0&0\\
0& 0  &0&0&0&0&0&0\\
0& 0  &0&0&0&0&0&0\\
0& 0  &0&0&0&0&0&0
\end{array} \right)\ 
\end{equation}
One immediately observes that all the broken gauge bosons are now massive,  while the unbroken gauge masses vanish, again as expected. Next, let us choose a basis where the mass matrix is diagonal:
\begin{equation}
\left( \begin{array}{c}
A_{10}'\\
A_{20}' 
\end{array} \right)\ \equiv 
\frac{1}{\sqrt{2}}\left( \begin{array}{cc}
1& -1\\
1& 1  
\end{array} \right)\ \left( \begin{array}{c}
A_{10}\\
A_{20} 
\end{array} \right)\ 
\end{equation}
The matrix now reads
\begin{equation}
M_{\as\bs}'=\frac{1200}{11}\left( \begin{array}{cccccccc}
\2& 0&0&0&0&0&0&0\\
0 & \2&0&0&0&0&0&0\\
0& 0  &1&0&0&0&0&0\\
0& 0  &0&1&0&0&0&0\\
0& 0  &0&0&1&0&0&0\\
0& 0  &0&0&0&0&0&0\\
0& 0  &0&0&0&0&0&0\\
0& 0  &0&0&0&0&0&0
\end{array} \right)\ 
\end{equation}
As is usual in Kaluza-Klein theories, the masses depend inversely on the ``radius'' $r$ (hidden in 3.13) of the compact internal manifold which can be seen as follows: 
$$y^{\ms}\ra ry^{\ms}\Ra T_{\as}\ra \frac{T_{\as}}{r}\Ra C_{\as\bs}{}^{\cs}\ra \frac{C_{\as\bs}{}^{\cs}}{r}\Ra M_{\as\bs}\ra \frac{M_{\as\bs}}{r^2}$$
Thus it is only the mass ratios of the gauge bosons that come out as a precise prediction of this model. In this case the mass squares  of $A_{0}$'s are half as those of $A_{A}$'s and the typical mass of the broken gauge bosons are given by
$$ M\sim \frac{10}{r}$$
{\bf Dynamical Considerations and Supergravity:}  We have seen  how in general geometric transition of the internal manifold can explain symmetry breaking of gauge theories. It is imperative then that we try to understand how or when such a transition can occur. If a classical path connects the two vacua, initial fluctuations and instabilities can then trigger a ``classical transition''. Alternatively, this could be brought about by  quantum tunneling effects. Although it might seem hopelessly difficult to analyse this ``infinite-mode'' coupled system, the ansatz for $g^S$ (2.14) gives us important clues as to what may be going on. The parameter $T^2$ which determines the ``shape'' of the manifold changes as it makes a transition. To maintain $\breve{\lambda}_{init}=\breve{\lambda}_{final}$, it is clear that the ``size'' of the internal manifold must also change. We also know how the external manifold (the four dimensional world) behaves asymptotically. For example, if we restrict ourselves to the case when $\Lambda >0$,  then we expect an asymptotically deSitter vacuum whose metric in a special coordinate system can be written as
$$ds^2=dt^2+e^{\frac{\Lambda t}{D}}dx^2$$
Thus it is natural to consider the full ansatz for the metric as
\begin{equation}
\hat{g}_{\ah\bh}= \left( \begin{array}{cccc}
-W^2(t) & 0       & 0 & 0\\
0  & A^2(t)\de_{ab} & 0 & 0\\
0  & 0       & S^2(t) g^K_{\az\bz} & 0\\
0  & 0       & 0                      &S^2(t) T^2(t) g^K_{\ad\bd}
\end{array} \right)
\end{equation}
$S(t)$ and $T(t)$ are to be treated as time dependent collective coordinates characterizing the size and the shape  of the manifold respectively, $A(t)$ is the usual cosmological radius of our universe while $W(t)$ corresponds to a gauge freedom (which may be useful for later computational purposes).
Asymptotically we expect $A(t) \sim e^{\frac{\Lambda t}{D}}$ (in the $W(t)=1$ gauge) before and after the transition, while $\{ S(t),T(t)\}$ should extrapolate between the two vacuum values. 

Classically, we now have to solve Einstein's equations (2.12). From the symmetry of (3.14) we expect to obtain four equations coming from the $\{tt,ab,\az\bz,\ad\bd\}$ component of (2.12), out of which one can be thought of as a gauge fixing condition. It should be pointed out that one may need to include matter/radiation contributions in Einstein's equations in a more realistic cosmoslogical setting and/or to incorporate SUGRA effects. At any case, in the end we will have four ordinary differential equations in four variables. However, such nonlinear differential equations coming from general relativity are usually notoriously difficult to solve and perhaps one has to employ both analytic and numerical techniques to analyse it in detail.

Alternatively, one can substitute the ansatz (3.14) in Einstein's action
\begin{equation} S_{D}=\int d^{D}x \sqrt{-\hat{g}}\hat{R} \end{equation}
and obtain an effective quantum mechanical action for $A(t),S(t)$ and $T(t)$\footnote{ Ideally one should check the consistency of the effective action with Einstein's equations.}.  It should then be possible to understand both qualitatively and quantitatively some of the quantum aspects of the problem. One can try to  compute the  quantum tunneling amplitude from the effective potential and also gain insight into a related and very crucial issue, that of stabilities of the vacua.

Stability of Kaluza-Klein vacuum has previously been studied \cite{stability}, specifically in the context of SUGRA theories. For example, it is well known that vacua's which preserve some supersymmetry are stable \cite{stability}, and thus it is important that we analyse the supersymmetric properties of the maximally symmetric and scaled metrics which are involved in the geometric transition\footnote{Although, in this paper I have focussed on pure gravity, it should  only be seen as a toy model to be ultimately embedded in a more unified (and presumably supersymmetric) theory of gravity like String and Supergravity theories.}. It should be mentioned that previously such ``less symmetric'' vacuum metrics on some coset spaces and their supersymmetric properties have been studied  in the context of supergravity compactifications and they go by the name of ``squashed spaces'' \cite{squashed}. Indeed some of these squashed spaces were found to admit Killing spinors, i.e. preserve some supersymmetry and hence are stable, making a geometric transition to these less symmetric spaces viable. Stable but  non-supersymmetric vacuum solutions also exist and their stability can be analysed by looking at the eigenvalues of the Lichnerowicz operator acting on some tensor fields on the internal manifold (please see \cite{horowitz} for some recent work). Similar methods can be adopted to study the vacuum solutions that I considered in this paper. 

In passing from pure gravity to supergravity we not only introduce fermions (and supersymmetry) but also usually some scalars and gauge forms. The presence of the gauge forms will enable us to consider not only the purely geometric vacua that we have been considering, but also the Freund-Rubin type vacua \cite{freund} with non-zero gauge form flux, and transitions between them. This is specially useful for SUGRA's with no cosmological constants to begin with (like the 11 dimensional SUGRA), because the stress-tensor of the gauge forms behave as an effective cosmological constant (except that it differs in sign for the external and the internal manifold) and thereby let us consider non-Ricci flat internal manifolds. In  Freund-Rubin type mechanisms the massless vectors recieve contributions (say $B_m^{\as}$) from the gauge forms and our analysis has to be suitably modified to incorporate this effect. However, presumably the geometric transition will preserve the masslessness of the same vector combinations (i.e. $A_m^{\ad}+B_m^{\ad}$) as in the pure gravity case, leaving the other combinations ($A_m^{\az}+B_m^{\az}$) broken. The fact that the gauge transformations (which follows from the general coordinate transformations in the extra dimensions, see the following section)of the $A_m^{\as}$'s both before and after the transition  remains unaltered even within the framework of SUGRA,  corroborates the above arguement. The $B_m^{\as}$'s have to essentially transform the same way as the  $A_m^{\as}$'s under the gauge symmetry $G$ and should also break the same way as the  $A_m^{\as}$'s under $H$, which suggests that the mass matrix should remain intact, barring an overall factor.

\section{ {\bf  HIGG'S LIKE MECHANISM }}
A natural question to ask at this point would be, what happens to the number of physical degrees of freedom! Clearly, for each broken generator, the gauge boson becomes massive from massless, thereby  increasing the number  of physical states by one. However, we will see that like in Higg's mechanism, here also the ``freed gauge parameter'' corresponding to the broken gauge bosons can be used to eliminate some of the scalars that are present in the full theory. To make this explicit let us identify, and include these scalars in the vielbein:
\begin{equation}
\eh_{\ah}{}^{\mh}=\left( \begin{array}{cc}
e_{a}{}^{m}(x) & \ e_a{}^n(x)A_n^{\bs}(x)\es_{\bs}{}^{\ms}(y)\\
0 &\ \Phi_{\as}{}^{\bs}(x)\es_{\bs}{}^{\ms}(y)
\end{array} \right)
\end{equation}
with
\begin{equation}
\Phi_{\as}{}^{\bs}(x)\equiv D_{\as}{}^{\bs}(\phi^{\cs}(x))=[exp(\phi^{\cs}(x)D_{\cs})]_{\as}{}^{\bs}
\end{equation}
where $D_{\cs}$ are the generators of $G$ in the adjoint representation. In other words, the scalars $\phi^{\cs}(x^m)$ are in the adjoint representation of the group $G$. 

Now let us look at the gauge transformations, which are essentially the translations along the Killing vectors of the internal manifold:
\begin{equation}
x^{'m}=x^m;\ x^{'\ms}=x^{\ms}+\ze^{\as}(x)\es_{\as}{}^{\ms}(y)
\end{equation}
Under this coordinate transformation, the gauge fields and the scalars transform as\footnote{for  infinitesimal $\ze$}
\begin{equation}
\de A_m^{\cs}=\p_m\ze^{\cs}(x)+C_{\as\bs}{}^{\cs}A_m^{\bs}\ze^{\as}
\end{equation}
and
\begin{equation}
\de \phi^{\as}=\ze^{\as}
\end{equation}
While (4.4) is the familiar nonabelian gauge transformation, using (4.5) we can eliminate the scalars that we introduced. We are now ready to count the number of states before and after the transition. The best way to do it is to look at the physical degrees of freedom associated with $\hat{g}_{\ms\ns}$ and $A_m^{\as}$. Now
$$
\hat{g}_{\ms\ns}=\eh_{\ms}{}^{\ah}\eh_{\ns}{}^{\bh}\hat{g}_{\ah\bh}=\eh_{\ms}{}^{\as}\eh_{\ns}{}^{\bs}\hat{g}^{\as\bs}=D_{\cs}{}^{\as}D_{\ds}{}^{\bs}\es_{\ms}{}^{\cs}\es_{\ns}{}^{\ds}\breve{g}_{\as\bs}$$
For the Killing metric we know from (9) $D_{\cs}{}^{\as}D_{\ds}{}^{\bs}\breve{g}_{\as\bs}^{K}=\breve{g}_{\ds\cs}^{K}$ and thus
\begin{equation}
\hat{g}_{\ms\ns}=\es_{\ms}{}^{\as}\es_{\ns}{}^{\bs}\breve{g}^K_{\as\bs}
\end{equation}
i.e. there are no  degrees of freedom associated with $\phi$'s to begin with\footnote{Note in general $\hat{g}_{\ms\ns}(x,y)$ can be expanded as a sum of harmonic functions/modes on the internal manifold. The coefficients are four-dimensional fields and constitutes the degrees of freedom associated with $\hat{g}_{\ms\ns}(x,y)$. (4.6) tells us that $\phi$'s are not associated with any such mode and hence do not constitute any degree of freedom in the theory.}. Like wise, the gauge parameters $\ze^{\as}$'s are used as gauge parameters for the massless $A^{\as}$ gauge fields. We thus have $2\breve{D}$ physical states coming from the gauge bosons, where $\breve{D}$ is the number of internal dimensions. 

For the scaled metric however, things are different. (2.9) tells us that 
$$D_{\cs}{}^{\as}(h)D_{\ds}{}^{\bs}(h)\breve{g}_{\as\bs}^{S}=\breve{g}_{\ds\cs}^{S} \mbox{ only when } h\ \e\ H$$
Thus we have 
\begin{equation}
\hat{g}_{\ms\ns}=(D_{\cs}{}^{\as}(\phi^{\az})D_{\ds}{}^{\bs}(\phi^{\az}))(x)(\breve{g}_{\as\bs}^S\es_{\ms}{}^{\cs}\es_{\ns}{}^{\ds})(y)
\end{equation}
and we see that although the $\phi^{\ad}$'s are spurious (as in the maximal case), the $\phi^{\az}$'s are associated with the mode functions $(\breve{g}_{\as\bs}^S\es_{\ms}{}^{\cs}\es_{\ns}{}^{\ds})(y)$, and hence are apriori present as genuine degrees of freedom in the theory for both the maximal and the non-maximal case. If we include these scalars, the total number of physical states in the maximal case goes up to
\begin{equation}
\mbox{physical states}_{maximal}=2\breve{D}+\grave{D}
\end{equation}
In the non-maximal case, we only need the $\ze^{\ad}$'s to preserve gauge symmetry, and the total number of physical states coming from the vectors are:
$$2\dot{D}+3\grave{D}=2\breve{D}+\grave{D}$$
The $\ze^{\az}$'s which are still left now can be utilized to eliminate the $\phi^{\az}$'s via (4.5). Thus the total number of physical states is again given by
\begin{equation}
\mbox{physical states}_{non-maximal}=2\breve{D}+\grave{D}
\end{equation}

Quite simply,  what is going on is that the number of degrees of freedom in the metric and the gauge parameter is the same for both the maximal and the non-maximal case. In the maximal case all the gauge parameters $\ze^{\as}$'s are utilized to eliminate degrees of freedom from the gauge fields, while in the non-maximal case the $\ze^{\ad}$'s do just that, but the $\ze^{\az}$'s eliminate some of the scalars. This is exactly analogous to what happens in Higg's mechanism. 

\section{{\bf SUMMARY AND FUTURE RESEARCH}} 

In this paper I have tried to explain how the dynamics of the internal manifold can break the gauge symmetry partially. As an illustrative example I considered the manifold $SU(3)$ which admits both a $SU(3)_L\otimes SU(3)_R$ and a  $SO(3)_L\otimes SU(3)_R$ invariant vacuum. If and when the metric makes a vacuum to vacuum transition from the maximal to the non-maximally symmetric space, the gauge fields corresponding to the broken Killing symmetries will pick up a mass, thereby breaking the gauge invariance from $SU(3)_L\otimes SU(3)_R$ to $SO(3)_L\otimes SU(3)_R$. This geometric symmetry breaking mechanism  has two definite predictions. First is a restriction of the subgroups into which the original gauge group can be broken, and this comes about because we cannot always find vacuum metrics invariant under a specific subgroup. Secondly, this model makes a precise prediction about the various mass ratios of the broken gauge bosons. The counting of physical states also works out fine when one properly takes into account some of the scalar degrees of freedom that are present in the full theory. The ``freed'' gauge parameter can eliminate these scalars in the non-maximal case, and thus the situation is exactly analogous to the Higg's mechanism where the gauge bosons become massive, ``eating away'' some of the scalars.

Though this geometric transition mechanism seems consistent, to make it more convincing one should study the transition itself. For example, why and how can such a transition occur? As a first test, we should  look for classical solutions connecting the two vacua (possibly asymptotically). If such a solution does not exist, quantum tunneling can still bring about such a transition. The stability of the scaled metric is of course crucial to gauruntee that the internal manifold remains there and hence has to be carefully analysed. Assuming that a geometric transition of the internal manifold is possible, our next task will be to understand its implications. I have already explained how this process will partially break gauge symmetry and thus  one can apply this procedure to break electro-weak symmetry or  grand unified theories (GUTs). The nice thing about this model is that it corroborates our general idea that the early universe was in a more symmetric phase, and then (through this geometric transition) came to exist in a less symmetric phase. To understand this theory in more detail, one will need to look at the truncation procedure carefully. Dynamical considerations suggest that some scalar fields (of comparable mass as the broken gauge bosons) play an important role in the effective theory, and their potential may be crucial in further understanding, like in the Higg's mechanism. Also, I performed the analysis only with pure Einstein-Hilbert action, and ignoring the dilaton, other gauge fields and the fermions. In a SUGRA theory, one obviously has to incorporate these fields and I have only hinted at what extensions/modifications that may be needed in our model. One should also investigate whether such a geometric transition has any  cosmological consequences.

Finally, it should be possible to generalise the internal manifolds and metrics that I considered in this paper. I analysed only the very simple case when the internal manifold is a Lie group, and when the vacuum metrics are all right invariant. One can try to look at other less symmetric metrics which are neither left nor right invariant. It should also not be too difficult to generalise this mechanism to more realistic internal manifolds like the coset spaces (and in particular those which can give Standard Model like gauge groups), and at least in principle to some of the more interesting nonhomogenous spaces. 
\vspace{5mm}\\
{\large {\bf Acknowledgements:}} I would like to thank Prof. Martin Rocek and Prof. Warren Siegel for some useful suggestions and discussions.

\end{document}